\documentclass[conference]{IEEEtran}
\usepackage{fancyhdr}
\IEEEoverridecommandlockouts

\usepackage{cite}
\usepackage{amsmath,amssymb,amsfonts}
\usepackage{graphicx}
\usepackage{textcomp}
\usepackage{xcolor}
\usepackage{tikz}
\usepackage{ctable}
\usepackage[ruled,linesnumbered]{algorithm2e}
\usepackage{xspace}
\usepackage{tcolorbox}
\usepackage{listings}

\def\BibTeX{{\rm B\kern-.05em{\sc i\kern-.025em b}\kern-.08em
T\kern-.1667em\lower.7ex\hbox{E}\kern-.125emX}}

\makeatletter
\def\lst@makecaption{
  \def\@captype{table}
  \@makecaption
}
\makeatother

\definecolor{myblue}{RGB}{63, 90, 126}
\definecolor{mygray}{RGB}{228, 244, 247}
\definecolor{pssgblue}{HTML}{0072B2}
\definecolor{pssgpurple}{HTML}{800080}
\definecolor{pssgdarkgreen}{HTML}{006B4F}

\newcounter{obsnum}

\begin{document}

\title{SpCCL: A Sparsity-Aware Collective Communication Library for GPU Platforms}

\author{
  \IEEEauthorblockN{Lannie Dalton Hough, Emir Gencer, Hoffmann Muki, Abhinav Bhatele}
  \IEEEauthorblockA{~\\
	  Department of Computer Science, University of Maryland, College Park, MD, USA\\
	  E-mail: \{ldhough, egencer, hoffmuki\}@umd.edu, bhatele@cs.umd.edu
  }
} 

\maketitle

\thispagestyle{fancy}
\lhead{}
\rhead{}
\chead{}
\lfoot{\footnotesize{
    SC26, November 15-20, 2026, Chicago, Illinois, USA
    \newline 979-8-3195-4789-7/26/\$31.00 \copyright 2026 IEEE}}
\rfoot{}
\cfoot{}
\renewcommand{\headrulewidth}{0pt}
\renewcommand{\footrulewidth}{0pt}

\begin{abstract}
Collective communication is essential to
high performance computing and machine learning workloads, yet
libraries such as NCCL do not exploit sparsity in message payloads.
Sending only nonzero values can reduce network traffic, but
explicitly handling sparsity introduces challenges such as
compression and decompression overheads.
We address these challenges with sparsity-exploiting versions of
all-gather, reduce-scatter, and all-reduce collectives.
Our implementations use a new bitvector-based format, Pici,
designed for low space overhead and fast GPU-based compression and decompression.
Further, our collective algorithms adapt to the degree of sparsity in data,
modifying data representations during the course of the collective. At 99\% input
sparsity, our collectives achieve up to 5.25$\times$, 2.5$\times$, and
2.66$\times$ speedups over NCCL for all-gather, reduce-scatter, and all-reduce,
respectively. Integrating our collectives into a representative
deep learning application, we achieve a 26\% end-to-end speedup.



\end{abstract}



\section{Introduction}

Sparsity,~i.e., the presence of many zero values
in data, is common in certain parallel workloads, including high performance
computing (HPC)~\cite{gpu_comm_conjugate, amesos2, superlu_sparse, qiu2024scalable},
and even more so in distributed deep learning (DL)~\cite{singh:ipdps2023,
lin2017deep, lottery, ranjan:sc2025}. In practice, the sparse data often
takes the form of an array or matrix of $N$ elements, of which $\mathit{nnz}$
are nonzero. The degree of sparsity is expressed by the fraction of
zeros as $\sigma = (N - \mathit{nnz}) / N$, often reported as a percentage.
When working with sparse data, it is often useful to quantify the
degree of sparsity with terms like ``highly sparse'' or ``low density.'' However,
such terms are not standardized, and what counts as highly sparse may
vary by problem domain or researcher. In this paper, we primarily contrast
two sparsity regimes. The high-sparsity (HS) regime comprises $>$99\% sparse data,
while the non-high-sparsity (NHS) regime comprises data with 1--99\% sparsity, inclusive.
We consider data that is less than 1\% sparse to be near-dense.

Exploiting sparsity in data (both
application input data and large arrays in memory) to improve the performance
of computation and communication is common~\cite{coruscant, sparcml,
omnireduce}. Communication optimizations center on sending only important
information (e.g.,~nonzero values), often in collective operations such as
all-gather, reduce-scatter, and all-reduce.  As communication overheads in
parallel applications can represent a significant fraction of execution time,
particularly as applications scale to more GPUs and nodes, such optimizations
can be particularly helpful for large messages.


Historically, sparsity-aware kernels and communication collectives have
often targeted the HS regime or specific
sparsity patterns. However, workloads such as the training of pruned deep neural networks
often exhibit unstructured sparsity patterns at lower degrees of
sparsity~\cite{singh:ipdps2023, lottery}. In this paper, we introduce
communication collectives that exploit unstructured sparsity across a wider
range of degrees of sparsity and significantly improve communication performance.


The task of implementing efficient sparsity-aware collectives (sparse
collectives, in short) is challenging because of several features of
problems with sparse data (sparse problems, in short):
\begin{itemize}
\item \textbf{Irregularity}: Unlike many dense problems, sparse problems are
highly variable in their characteristics. For example, both the degree of
sparsity and the sparsity structure
(where the nonzeros occur) can vary and impact which
algorithms, kernels, and sparse formats are effective.
\item \textbf{Dynamic changes}: In the case of collectives that include a
reduction such as reduce-scatter and all-reduce, the degree of sparsity can
        decrease across reduction steps during the collective (``dynamic sparsity'').
        If processes contribute sparse data with
differing nonzero indices, the sparsity of intermediate and final outputs
decreases (``fill-in'').  Thus, better support for non-high-sparsity data is
required in such cases.
\item \textbf{Overheads}: Conversion between dense and sparse formats
(compression and decompression) can incur significant overheads. We must
ensure that these overheads do not offset the benefits of communicating
less data.
\end{itemize}

Additionally, even when data has a large number of zeros, operations surrounding
the collective may still assume a dense representation. These concerns
indicate that we must consider not only how space-efficient a sparse format is
but also the performance of its compression and decompression kernels.


We observe that traditional formats for representing sparse data, such as Coordinate
(COO)~\cite{bell2008efficient} and Compressed Sparse
Row (CSR)~\cite{eisenstat1977yale}, are poorly suited to collectives
where fill-in quickly drives data outside of the HS regime. 
These formats use a large amount of metadata per nonzero element. While
acceptable for data in the HS regime, this becomes increasingly undesirable
as sparsity decreases.
We introduce a novel bitvector-based format, Pici, which instead uses a
small constant amount of metadata per input element.

We also observe that below a certain sparsity threshold, which
depends on many factors, including compression and decompression speeds,
compression and decompression frequency, and the bandwidth of the communication
path, the overhead of
compression and decompression outweighs the benefits of communicating less data. Due to
fill-in in collectives with reductions, this may occur after some intermediate
steps in the collective algorithm that do benefit from exploiting sparsity initially. To
address this concern, we design our collectives to adapt to dynamic sparsity,
changing the representation of data from sparse to dense as needed during the course of the collective.
To leverage prior collective optimization work, we implement our sparse
collectives within a fork of the NCCLX collective communication
library~\cite{si2025collective} and name it SpCCL (\textbf{Sp}arsity-Aware
\textbf{C}ollective \textbf{C}ommunication \textbf{L}ibrary).


We evaluate SpCCL's collectives on NERSC's flagship Perlmutter supercomputer
\cite{perlmutter}. For a range of input sparsities, we achieve improved
performance compared to both other sparse collective implementations and 
NCCL~\cite{nccl} and NCCLX~\cite{si2025collective}. We also incorporate our collectives into a
representative deep learning (DL) application and demonstrate end-to-end
performance speedups.


This paper makes the following key contributions:
\begin{itemize}
  \item We introduce Pici (pronounced ``peachy''), a novel bitvector-based
sparse representation that is space-efficient across a wide range of sparsity levels,
        along with supporting CUDA kernels. This
format imposes a low, sparsity-independent metadata overhead of approximately $3.15\%$ for fp32 data,
while also supporting efficient parallel compression and decompression.
  \item We develop optimized implementations of three sparse
collectives that use Pici: all-gather, reduce-scatter, and all-reduce. We focus on effective
support for unstructured data in the NHS regime, but SpCCL collectives are also effective
in the HS regime.
  \item We mitigate the problem of fill-in by introducing a sparsity-aware
  algorithm that adapts the data representation depending on the degree of
  sparsity in the data, the current stage of the collective, and the links that
  messages traverse.
  \item Using SpCCL with Pici-based collectives, we achieve substantial speedups over
state-of-the-art dense communication libraries and a prior sparsity-aware all-reduce implementation.
        We also demonstrate
gains in a representative DL application. At 99\% sparsity, SpCCL outperforms a
dense NCCL baseline for all-gather, reduce-scatter, and all-reduce by up to
5.25$\times$, 2.5$\times$, and 2.66$\times$, respectively.
\end{itemize}

\section{Background}\label{background-section}
In this section, we provide the background necessary to understand requirements
and tradeoffs involved in representing sparse data and implementing sparse
collectives.

\subsection{Modeling Performance of Collective Routines}\label{background-coll-subsection}

Our work focuses on three collectives: all-gather (AG), reduce-scatter (RS),
and all-reduce (AR).
Libraries frequently implement all-reduce as RS+AG
\cite{rabenseifner2004optimization}.  Before designing a sparse collective, we
consider a cost model to determine under what conditions we can realize
speedups. The Hockney model is a simple and effective way to model
communication \cite{hockney1994communication} and is sufficient to reason about
the ideas this paper introduces, although more sophisticated models do exist
(such as LogGP~\cite{loggp} and max-rate~\cite{maxrate}).  Under the Hockney
model, the $\alpha$ term represents a message's startup latency.
The $\beta$ term represents the inverse of a link's peer-to-peer bandwidth.
$n$ describes the message size in bytes. Taken together:
\begin{equation}
  T_{\mathrm{comm}} = \alpha+n\beta
\end{equation}

We use the Hockney model to characterize how different collective communication
algorithms affect each term. For example, with the ring algorithm, each process
communicates with immediate neighbors in a circular fashion, forwarding
messages around the ring in $p-1$ steps. With $n$ as per-process output payload,
ring all-gather can be modeled as:
\begin{equation}
    T_{\mathrm{ring}} = (p-1) \times \left ( \alpha + \frac{n}{p} \times \beta \right )
\end{equation}

While bandwidth-optimal for dense payloads, ring can suffer at very large process counts due to
the latency term, which scales linearly. In contrast, some recursive algorithms
divide the communication task into $\log_2(p)$ steps, either doubling (AG) or
halving (RS) the message volume at each step.

Compression yields speedups by reducing transmitted message volume. This implies that
if messages are already small enough for $\alpha$ to be dominant, further
reducing communication volume is unlikely to achieve significant speedups. As a
consequence, we focus on the large message regime (densely formatted inputs of hundreds of
MiB).  Large messages are typical of distributed training, for example, when
synchronizing gradients using distributed data parallelism (DDP)
\cite{pytorchdist-vldb}.

Although the Hockney model is useful for modeling dense collectives, it is
insufficient for sparse collectives, where compression and decompression introduce additional
overhead.
We extend the model with two additional terms, $T_{\mathrm{comp}}$ and $T_{\mathrm{decomp}}$:
\begin{equation}
  T_{\mathrm{comm\_sparse}} = \alpha+n\beta + T_{\mathrm{comp}} + T_{\mathrm{decomp}}
\end{equation}

The impact of these additional terms varies depending on the collective in
question. Additionally, $n$ in $T_{\mathrm{comm\_sparse}}$ represents the sparse payload
of the modeled message, which depends on format choices and may vary across
algorithm steps. While we defer a more detailed discussion of these
considerations to Section \ref{ccd-section}, minimizing both $T_{\mathrm{comp}}$ and
$T_{\mathrm{decomp}}$ is clearly beneficial. This necessitates a sparse format with
high-performance compression and decompression kernels.

\subsection{Overview of Sparse Data Formats}\label{sparse-format-section}

Existing literature describes numerous sparse representations
\cite{langr2015evaluation,gao2023systematic}. In order to reconstruct (decompress)
densely formatted data, sparse formats should store nonzero values and metadata that
can be used to identify the positions of the nonzero values.
Two of the most common and
general-purpose formats are Coordinate (COO) and Compressed Sparse Row (CSR)
\cite{bell2008efficient,eisenstat1977yale}. COO represents sparse matrices by
storing row and column indices for each nonzero in the matrix, as well as the
value itself. CSR uses less metadata than conventional COO by omitting
per-nonzero row indices and instead storing a per-row prefix sum of the number
of nonzeros ($\mathit{nnz}$) encountered. This generally results in more complex
compression and decompression routines.

A common variant of COO treats inputs as one-dimensional, storing only a single
flat index per nonzero. This cuts index metadata in half.
For the rest of
this work, COO refers to this variant. Prior work uses COO in sparse
collectives, such as SparCML's all-reduce implementations \cite{sparcml}.
Alongside Pici, we use COO in this work. Space overhead
calculations assume 32-bit indices and 32-bit values.

Observe that in
``index + value'' formats, such as COO and CSR, metadata overhead scales
linearly with $\mathit{nnz}$.  Relative to formats that impose a
sparsity-independent overhead
of $m$ metadata bytes, COO uses less space
when $4 \times \mathit{nnz} < m$. While often requiring less metadata than
alternatives at higher sparsities,
COO becomes less competitive with
formats featuring sparsity-independent metadata overhead as sparsity decreases.
At $50\%$ sparsity, COO requires the same amount of space
as the dense representation, and so
space savings cannot be realized.

Other formats exploit structure in sparse data \cite{saad1990sparskit}.
Block Sparse Row (BSR)
generalizes CSR to \textit{blocks} of nonzeros and explicitly stores zeros
within a nonzero block. Diagonal (DIA) stores nonzero diagonals. Structured
formats such as BSR can be highly space-efficient when the sparsity pattern
closely matches the structure being exploited by the format.
However, structured formats can be highly inefficient for
unstructured sparse data. Consider BSR with $16 \times 16$ blocks, where each
element has an independent 99\% chance of being zero. The chance of an entire
block being zero is only $0.99^{256} \approx 0.076 = 7.6\%$, and all of the
zero values in a nonzero block must still be stored.

Finally,
bitvector-based formats use bitvectors to implicitly encode the positions of
nonzeros. The simplest variant of this is Bit-Vector (BV), which stores a
single bit per input element and each nonzero
value in a compacted values array \cite{kestur2012towards}. Bits are set to
1 for corresponding nonzero elements or 0 otherwise.
Compressed Bit-Vector (CBV) reduces space by encoding runs of zeros and nonzeros.
Observe that for BV, metadata size does not depend on the sparsity pattern.
Its low, sparsity-independent metadata
overhead makes BV appealing across much of the non-high-sparsity regime.
For 32-bit values, BV has $3.125\%$ metadata overhead relative to the
dense representation. Thus, below
$96.875\%$ sparsity, BV is more space-efficient than COO.
For a comprehensive overview of sparse formats, we refer the reader to Langr et al.
\cite{langr2015evaluation}.

\section{Pici: A Space-Efficient Sparse Representation}
We now describe Pici, the design of its compression and decompression
algorithms, and the implementation of corresponding CUDA kernels.

\subsection{Key Design Considerations for Pici}

A sparse representation suitable for supporting arbitrary sparsity
patterns for data in the NHS regime and for mitigating the impact of the fill-in problem
should~(1)~have low space overhead for
data of varying degrees of sparsity, and~(2)~support efficient parallel
compression and decompression. Recall that fill-in is the densification
of data across reduction steps in multi-stage reducing collectives.

Among the formats described in Section \ref{sparse-format-section},
Bit-Vector (BV) satisfies the first design constraint particularly well. BV's metadata
overhead is independent of how nonzeros are distributed, so it is
suitable for unstructured sparse data. The overhead is also
sparsity-independent, and as a result, it does not increase
as sparsity decreases due to fill-in. Hence, 
we use a bitvector-based sparse representation as the foundation for designing Pici.

Unfortunately, raw bitvectors are not ideal for satisfying the second
design requirement. Consider parallel decompression with an
$N$-element bitvector $B$ and an array of nonzero values $A$ (the components
of the BV format). Na\"ively, threads start scanning
from different segments of the bitvector in order to parallelize scans that
identify the positions of nonzero values.
For a thread starting the scan at an arbitrary position $i$, if
$B[i] = 1$, the algorithm knows that a nonzero value should be placed at
position $i$ in the decompressed dense buffer. However, unless $i = 0$,
the algorithm does not know which value from $A$ corresponds to this
position. The algorithm needs to know how many bits in $B$ are set to 1
before position $i$.

Given an $N$-element bitvector, we could solve the value-lookup
problem with access to a metadata array $I$ storing, for each bit, the
number of preceding 1-bits in the bitvector (a prefix sum).
This na\"ive approach violates the low space overhead requirement because
of the high storage requirements of $I$. In practice,
bitvector-based formats may mitigate this problem by dividing the bitvector
into spans of length $k$ and storing only $\lceil N/k\rceil$ prefix sums. A thread
can then initiate a $k$-element scan from the start of any span.
With one thread available per span, each thread scans at most $k$ elements.

Observe that increasing $k$ reduces index overhead but decreases the number of
immediate offsets into $A$. This saves space but reduces the
number of spans that can be decompressed independently without additional
prefix computation.

To satisfy our design requirements, we ideally want to keep index overhead low
while facilitating high
parallelism. To achieve this, we take inspiration from the field of succinct data structures.
Works in this area demonstrate that in exchange for a modest amount of
span-level computation, data structures that can answer $\operatorname{rank}$ queries on bitvectors
may achieve significant index space savings~\cite{zhou2013space,kurpicz2022engineering}.
Informally, rank denotes the number of 1s or
0s in a bitvector, up to some position $i$. Answering $\operatorname{rank}_1$ queries is
isomorphic to our problem of determining the nonzero count up to the start of an arbitrary span.

Rank-enabling data structures such as poppy and pasta introduce many ideas to
keep space overhead and query time low
\cite{zhou2013space,kurpicz2022engineering}. They feature hierarchical indices
of decreasing width covering increasingly small spans
of the bitvector. The indices
are interleaved to minimize cache misses. For their
most granular indexed span of 512 bits, the \texttt{POPCNT} instruction enables faster scans,
as it can return the number of 1-bits in a 64-bit word while taking only a few
cycles.

The space overhead of poppy and pasta is exceptionally low, at 3.125\% relative
to the bitvector. However, these structures are still not ideal for supporting parallel
compression and decompression on GPUs. They target low-latency random queries on CPUs,
but within a thread, compression and decompression scan many adjacent bits or elements in series.
Rather than adopting these structures directly, we leverage the key insight that modest
span-local computation can reduce index space and design a data structure to
better fit our access patterns and the CUDA programming model.

\subsection{Description of the Pici Sparse Data Format}

The Pici data format consists of three parts: two metadata arrays and an
array $A$ of nonzero values from the dense representation of the original data. The
first metadata array is a bitvector, $B$, used to identify nonzero
positions, similar to the BV format.  The second metadata array is an index
array, $I$, which stores prefix sums of nonzero counts to provide immediate
offsets into the values array, $A$.

\begin{figure}[h]
  \centering
  \includegraphics[width=\linewidth]{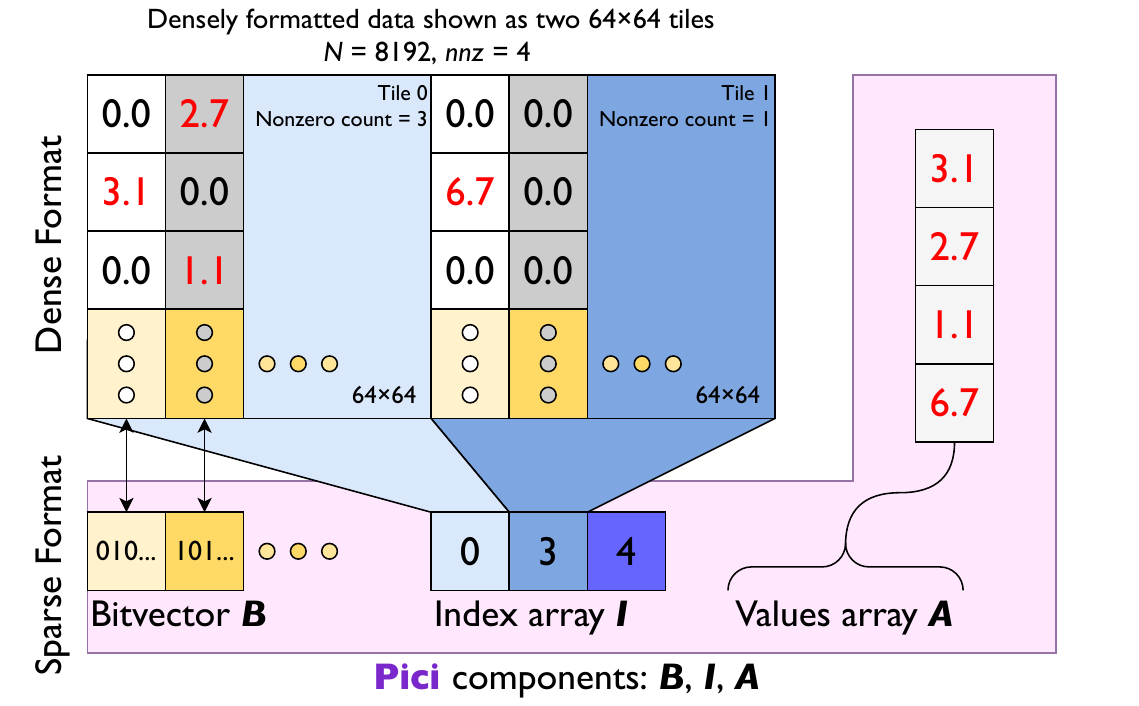}
  \caption{Overview of how a densely formatted input of $N=8192$ elements covered
    by two virtual tiles is represented in the Pici sparse format.}
  \label{fig:spop_design}
\end{figure}

To construct the three Pici components, the compression process receives
an $N$-element densely formatted input (a contiguous buffer of zero and nonzero elements).
Pici then views the densely formatted input as being composed of $64 \times 64$ row-major
``virtual'' tiles (see Figure
\ref{fig:spop_design}). Each tile corresponds to a 4096-bit span of $B$ and one exclusive
prefix sum in $I$. Because $I$ contains one prefix sum per 4096-bit span of $B$, $k=4096$.
A single additional entry at the end of $I$ stores the total number of nonzeros ($\mathit{nnz}$) in
the input.

Although the layout of tile elements is row-major,
each 64-bit word in $B$ corresponds to a single tile \textit{column}.
Because of Pici's tile- and column-based view into the input,
the bitvectors of Pici and BV are generally not identical.
The layout of nonzeros in Pici's $A$ also follows the view hierarchy, and
compacted nonzeros are ordered first by their parent tile, next by their
parent column, and finally by their intra-column position.
The choices of $k=4096$ and the column-first tile view are informed by
the CUDA programming model to facilitate fast compression and decompression,
discussed in more detail in Section \ref{gpu-kernels}.

Observe that while Pici features supporting indices in $I$,
it uses a flat indexing scheme instead of poppy-style hierarchical and packed
indices, avoiding overhead from packing variable-width indices.
Additionally, with a substantially larger $k$ than
poppy and pasta, Pici's index overhead is even lower, at only $\approx0.78\%$ relative to the
bitvector. Pici's total overhead relative to a 32-bit dense representation is only $\approx3.15\%$. Figure
\ref{fig:spop_design} shows how Pici components map to the tile-based view of
the densely formatted source.

\subsection{GPU Kernels for Compression and Decompression}\label{gpu-kernels}

Next, we describe the implementation of high-performance compression and decompression
kernels in CUDA. We begin with compression.
Data in Pici tiles is accessed by GPU threads in a warp. In CUDA, a ``warp'' is
a group of threads that executes instructions using the SIMT model.  When a warp is
processing data, the ideal access pattern is for threads to access adjacent
locations in memory. These accesses are referred to as ``coalesced'' memory
accesses.  For compression, we assign each tile to be processed cooperatively
by a single warp of 32 threads, such that each thread is responsible for two
columns. Now our decision to organize row-major tiles into columns makes sense.
As threads check elements in adjacent columns for nonzeros, their memory accesses
are coalesced.
If there are more tiles than available warps, warps
process multiple tiles. Algorithm \ref{algo:pici-compression} presents a
high-level, simplified view of compression, which proceeds in three phases.

\begin{algorithm}[h]
  {\footnotesize
    \KwIn{Densely formatted input array $D$ of length $N$}
    \KwOut{Bitvector $B$, index array $I$, values array $A$}
    \tcp{\textcolor{pssgblue}{Phase 1: Populate $B$, partial $I$}}
    \ForEach{\rm $tile\_idx$ assigned to warp \textbf{in parallel} across warps}{
        $t \gets \textsc{TileView}(D, tile\_idx)$\;
        \ForEach{\rm $lane\_idx \in \{0,\ldots,31\}$ \textbf{in parallel} across threads}{
            \tcp{Assigned cols are strided by 32}
            $(c\_1,c\_2) \gets (lane\_idx,lane\_idx+32)$\;
            \tcp{(col nnz, local 64-bit col bitvector)}
            $(nnz\_c1, bv\_c1) \gets \textsc{ScanColumn}(t, c\_1)$\label{line:pici-scan-first}\;
            $(nnz\_c2, bv\_c2) \gets \textsc{ScanColumn}(t, c\_2)$\label{line:pici-scan-second}\;
            $B[tile\_idx \times 64 + lane\_idx] \gets bv\_c1$\label{line:pici-bv-first}\;
            $B[tile\_idx \times 64 + lane\_idx + 32] \gets bv\_c2$\label{line:pici-bv-second}\;
            $nnz\_tile \gets \textsc{WarpShuffleReduce}(nnz\_c1 + nnz\_c2)$\label{line:pici-shuffle-reduce}\;
            \If{$lane\_idx = 0$}{
                $I[tile\_idx] \gets nnz\_tile$\label{line:pici-store-tile-count}\;
            }
        }
    }
    \tcp{\textcolor{pssgpurple}{Phase 2: Complete $I$}}
    $\textsc{SyncWorkers}()$\label{line:pici-phase2-start}\;
    $\textsc{ExclusivePrefixSumAndStoreTotalNNZ}(I)$\;
    $\textsc{SyncWorkers}()$\label{line:pici-phase2-end}\;
    \tcp{\textcolor{pssgdarkgreen}{Phase 3: Compact values array $A$}}
    \ForEach{\rm $tile\_idx$ assigned to warp \textbf{in parallel} across warps}{
        $t \gets \textsc{TileView}(D, tile\_idx)$\;
        \ForEach{\rm $lane\_idx \in \{0,\ldots,31\}$ \textbf{in parallel} across threads}{
            \tcp{Thread-assigned cols are adjacent}
            $(c\_1,c\_2) \gets (2 \times lane\_idx,2 \times lane\_idx+1)$\;
            $bv\_c1 \gets B[tile\_idx \times 64+c\_1]$\;
            $bv\_c2 \gets B[tile\_idx \times 64+c\_2]$\;
            $(nnz\_c1,nnz\_c2) \gets \textsc{Popcounts}(bv\_c1, bv\_c2)$\label{line:pici-popcounts}\;
            $pair\_off \gets \textsc{WarpExPrefixSum}(nnz\_c1+nnz\_c2)$\label{line:pici-column-prefix}\;
            $offset\_c1 \gets I[tile\_idx]+pair\_off$\label{line:pici-offset-first}\;
            $offset\_c2 \gets offset\_c1+nnz\_c1$\label{line:pici-offset-second}\;
            $\textsc{CompactColumnIntoA}(t,c\_1,bv\_c1,A,offset\_c1)$\label{line:pici-compact-first}\;
            $\textsc{CompactColumnIntoA}(t,c\_2,bv\_c2,A,offset\_c2)$\label{line:pici-compact-second}\;
        }
    }

  }
  \caption{High-level, simplified Pici compression}
  \label{algo:pici-compression}
\end{algorithm}

In \textcolor{pssgblue}{\textbf{Phase 1}} of Algorithm~\ref{algo:pici-compression},
each thread in the warp scans the two columns it is
responsible for (Lines~\ref{line:pici-scan-first}--\ref{line:pici-scan-second}).
When encountering nonzero values, threads flip a bit in a
local word corresponding to the column and increment local column counters.
The kernel batches writes to $B$ in 64-bit words, avoiding per-bitflip writes
(Lines~\ref{line:pici-bv-first}--\ref{line:pici-bv-second}).
To obtain the full tile
nonzero count, the warp reduces the local column counts in parallel using the
\texttt{\_\_shfl\_down\_sync} warp-shuffle intrinsic
(Line~\ref{line:pici-shuffle-reduce}).
CUDA warp shuffles allow
for the direct register-to-register exchange of values across a warp's threads,
bypassing slower DRAM and shared memory. The reduction requires only $\log_2
32=5$ shuffles, as one reduction step uses the two thread-local column counts.
After reduction, the first lane in the warp writes the tile count to the
index array $I$ (Line~\ref{line:pici-store-tile-count}).

In \textcolor{pssgpurple}{\textbf{Phase 2}} of Algorithm~\ref{algo:pici-compression},
we prefix-sum the tile counts to complete the prefix sums in $I$
(Lines~\ref{line:pici-phase2-start}--\ref{line:pici-phase2-end}).
Typically, this operation would require synchronization across CUDA
blocks/streaming multiprocessors (SMs). However, in SpCCL, we ultimately
compress and decompress independent data segments within each CUDA block used
by collectives. We discuss the segmentation regime in more detail in Section
\ref{ncclx-section}. The segmentation allows us to instead implement a simple
block-local
prefix sum and fuse phases 1-3 together in a single device function.

In \textcolor{pssgdarkgreen}{\textbf{Phase 3}} of Algorithm~\ref{algo:pici-compression}, we compact nonzero values into $A$ using the now fully constructed
$B$ and $I$. Threads still process two columns in each tile (two batches
of columns per tile). Except for column zero, the number of nonzeros preceding each column is
not immediately available. We build per-column offsets into $A$ on the fly
using a Hillis-Steele-style prefix sum implemented via warp shuffles on
per-column nonzero counts obtained by the \texttt{\_\_popcll} popcount intrinsic (two
per thread, Lines~\ref{line:pici-popcounts}--\ref{line:pici-offset-second})
\cite{HillisSteele1986}. To quickly identify nonzeros
within a column, we apply to bitvector words a combination of masking and the
\texttt{\_\_ffsll} intrinsic, which identifies the location of the word's least
significant 1-bit. These operations occur within the
\textsc{CompactColumnIntoA} calls
on Lines~\ref{line:pici-compact-first}--\ref{line:pici-compact-second}.
As per-column nonzero counts may vary, warp divergence can
occur here, but the maximum per-column nonzero count in the column batch bounds loop
iterations. Depending on data sparsity, the number of iterations may be
significantly fewer than 64.

Decompression follows essentially the same procedure as Phase 3 of compression
but moves values from the compacted values array into a dense buffer. We also
implement a scatter-add variant of ``decompression,'' facilitating reductions
in all-reduce and reduce-scatter. Note how the popcount intrinsic and warp shuffles
allow Pici to use a relatively large $k$ value of 4096 while still supporting 
intra-tile parallelism. Using popcounts
executed in parallel across a warp's threads, followed by a short warp-shuffle prefix
sum, we can produce temporary prefix sums for each column in a tile. This lets Pici
retain the low index overhead of $k=4096$ while still efficiently computing temporary
column-granularity offsets into $A$ and using them to distribute a tile's work across 32 threads.

\section{Sparse Collective Algorithms in SpCCL}\label{ccd-section}
Next, we describe the design of our collectives. We focus on the
algorithmic and format-selection decisions that we make, deferring
implementation details to Section \ref{ncclx-section}.

\subsection{Selection of Collective Algorithms}

We must select a base algorithm for each of our communication collectives. For
example, we could use a ring algorithm or recursive doubling/halving as
discussed in Section \ref{background-coll-subsection}. However, we are limited
to algorithms supported by the NCCL-derived NCCLX path that SpCCL
extends.
For reduce-scatter and all-gather, only Ring is available on Perlmutter.
Some algorithms such as CollNet require
specialized hardware, and NCCLX only implements Parallel Aggregated Trees for
inter-node communication \cite{jeaugey2025patnewalgorithmallgather}. For
all-reduce, NCCLX also implements and supports the Tree algorithm on
Perlmutter. Similar to the recursive algorithms described in Section
\ref{background-coll-subsection}, Tree has a logarithmic number of
communication steps.

To choose between Ring and Tree for all-reduce, we consider empirical results
from several collective communication studies
\cite{huang2021communication,thakur2005optimization,singh2026pccl}. These studies find
that for large messages, Ring often performs better empirically than recursive
algorithms, despite having a worse latency term and communicating the same data
volume.  The authors attribute this to the ring communication pattern, which
can support effective pipelining and bandwidth utilization. We validate
this intuition by comparing NCCLX's Ring and Tree algorithms for a large
collective input of 512 MiB and a smaller input of 50 MiB.
Ring outperforms Tree at both sizes, suggesting
that it is an appropriate choice for our regime. All of our sparse collective
implementations utilize Ring.

\subsection{Where and When SpCCL Compresses and Decompresses}

Compression and decompression are necessary for sparse communication to occur,
but their placement and frequency within the algorithm differ by
collective. All-gather is the simplest and most forgiving case.  Data is
unmodified during the course of the collective, so compression can occur once
before communication. In our implementation, each process compresses its
contribution to the output before the first ring communication step and
forwards the sparse representation around the ring unmodified. As a process
receives shards of data from peers at each step, it immediately decompresses
them into the output buffer. Consider an all-gather on $p$ processes with an
expected output buffer of size $n$, data sparsity $\sigma$, and
$\omega=0.0315$ representing Pici's metadata overhead. Our extended, idealized cost model is:
\begin{equation}
    n' = n \times (1 - \sigma + \omega) \\
\end{equation}
\begin{equation}
    T_{\mathrm{cd}} = T_{\mathrm{comp}}(\frac{n}{p}) + T_{\mathrm{decomp}}(n' \times \frac{p-1}{p})
\end{equation}
\begin{equation}
    T_{\mathrm{sparse\_ring}} = (p-1) \times \left ( \alpha + \beta \times \frac{n'}{p} \right ) + T_{\mathrm{cd}}
\end{equation}

On the other hand, reduce-scatter must accumulate data after receipt from peers
at each ring step. We accumulate into a dense buffer using our Pici scatter-add
kernel. For the next communication step to benefit from sparsity, this
accumulated dense buffer must be recompressed. Thus, unlike all-gather,
reduce-scatter must compress data at each of the $p-1$ ring steps. Because
reduce-scatter has additional computational overhead and densifies data, we
expect smaller speedups for reduce-scatter than for all-gather. Our ring
all-reduce is implemented as a two-phase RS+AG, and each phase
mirrors the behavior of our standalone RS/AG collectives.

\subsection{An Adaptive Algorithm for Dynamic Sparsity}

One purpose of a sparse representation such as COO or
Pici is to reduce storage requirements and communication volume
relative to a dense representation. However, metadata
requirements result in the size of the
sparse representation matching or exceeding the dense one for low sparsities (for example, at 50\%
sparsity with COO). In these cases, communicating the sparse
representation would be wasteful,
especially when compression and decompression overheads are considered. A simple
heuristic for communication collectives is to swap data to a dense
format when the size of the sparse representation exceeds the dense
representation in size.
This
observation has been made previously in other work, such as SparCML
(using COO), which uses it to swap to a dense format during all-reduce
reduction steps \cite{sparcml}.

However, the format-size heuristic does not adequately account for the
holistic performance of sparse collectives. Swapping formats (whether COO or
Pici) based only on representation sizes is insufficiently
conservative unless $T_{\mathrm{comp}} = T_{\mathrm{decomp}} \approx 0$, which is unrealistic.

For our collectives, we add an adjustable threshold $\tau$ that switches to
communicating a dense representation when sparsity $\sigma$ is $\leq \tau$.
This decision is typically informed
by the degree of sparsity computed as a byproduct of an earlier
compression. However, the reduction process does not naturally compute the
post-reduction sparsity after receiving a peer's data, so the decision of
whether to compress data before sending is made based on an
\textit{extrapolated} post-reduction sparsity. We assume uniform sparsity and
independent positions across processes and extrapolate according to the following
formula, derived from the inclusion-exclusion principle:
\begin{equation}
    \sigma_{\text{next}} = \sigma_{\text{prev}}\sigma_0
\label{fig:extrapolate_equation}
\end{equation}
where $\sigma_{\text{next}}$ is the extrapolated sparsity,
$\sigma_{\text{prev}}$ is the previously computed sparsity, and $\sigma_0$ is
the initially computed sparsity, measured at step 0.  Step 0 always executes
compression, and the first threshold-informed send decision is based directly
on the computed sparsity.
For example, if we set the threshold to $0.6$ and sparsity is extrapolated to
be $\leq$60\% during an intermediate step, the collective
sends densely formatted data.

We also consider that in modern GPU clusters, communication typically occurs
over heterogeneous networks. For example, Perlmutter nodes feature four 25
GB/s/direction NVLink links connecting each GPU pair, while each GPU node is
equipped with four 25 GB/s/direction Slingshot 11 NICs. Thus, nominal peak
aggregate off-node injection bandwidth is equivalent to the one-way bandwidth
between a single pair of GPUs.
Consequently, inter-node sparse sends should more readily provide speedups.
To exploit this, we introduce \textbf{link-aware}
adaptivity, where the collective makes decisions on send format based on
whether it is sending data over inter- or intra-node links.  We support this
with a second threshold: $intra\_thresh$.  This threshold controls intra-node
send decisions, while our first threshold, now $inter\_thresh$, controls
inter-node decisions. These are only relevant to reduce-scatter and all-reduce,
as all-gather does not densify data.

\begin{figure}[h]
  \centering
  \includegraphics[width=\linewidth]{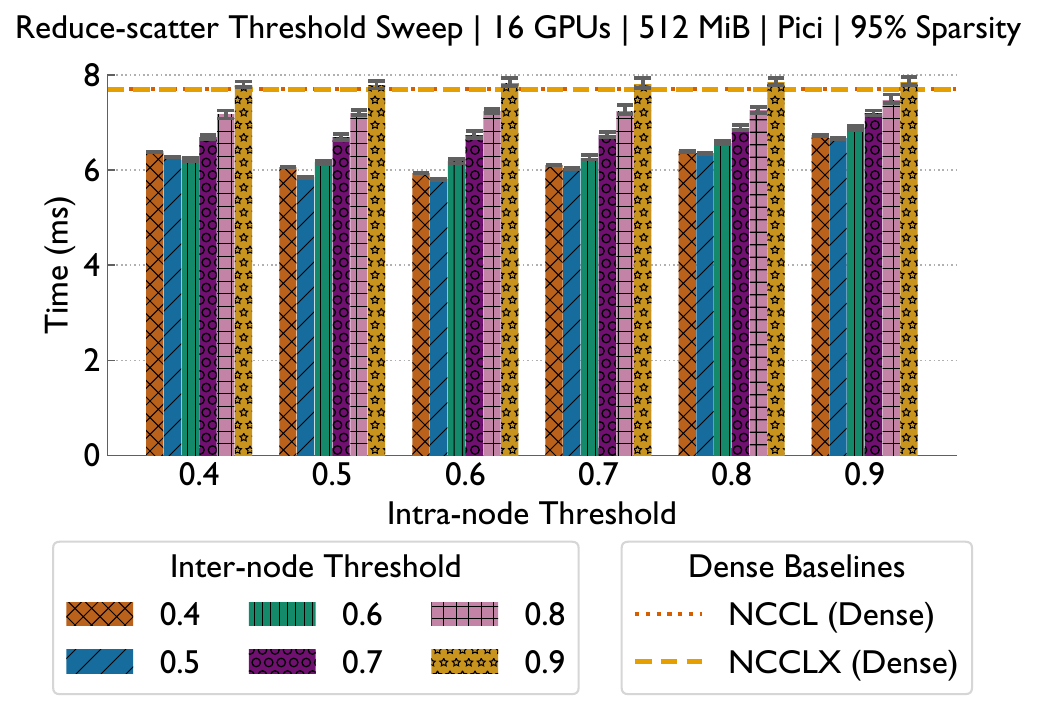}
    \caption{Impact of intra- and inter-node threshold values on reduce-scatter performance with Pici using a 512 MiB input size.
    Initial input sparsity is 95\%, resulting in
    substantial densification on 16 GPUs. Observe that swapping to dense too soon (high thresholds) or too late (low thresholds) limits the performance
    improvements that can be realized over dense.}
  \label{fig:rs_threshold_slice}
\end{figure}

Notably, $inter\_thresh$ and $intra\_thresh$ are not independent. Consider a
process that will post an inter-node send after receiving a sparse representation from an
intra-node send. It must scatter-add this data into a dense buffer, requiring
a more irregular memory access pattern.
Thus, the decision made at the prior intra-node send step affects the runtime
of the inter-node send step. To identify appropriate thresholds, we sweep
combinations of threshold pairs at intervals of $0.1$ across various GPU counts
and input/output buffer sizes for SpCCL using Pici.  Figure \ref{fig:rs_threshold_slice}
illustrates a subset of one of these sweeps. We observe that for both
thresholds, thresholding either too aggressively or too conservatively results
in slowdowns.  While the exact best thresholds are mildly sensitive to
collective input size, we find that for fp32 reduce-scatter,
$intra\_thresh=0.6$ and $inter\_thresh=0.5$ are generally strong choices.

Finally, for all-reduce, the expected ideal threshold varies by algorithm phase
(RS/AG). If, as suggested by our cost model, the sparse all-gather phase can
achieve speedups at lower sparsities, it does not make sense to forward data in a dense
format through all-gather just because the reduce-scatter phase terminated with
a dense representation. We propose \textbf{phase-aware} adaptivity and introduce an
all-gather threshold: $ag\_thresh$. This determines at the RS-AG transition
step in all-reduce whether the AG phase should proceed with a sparse or dense
representation. We find that a value of $0.1$ is appropriate for all-gather, which is, as
expected, much lower than the RS thresholds.

Although ideally we could compute thresholds analytically, the thresholds
depend on too many interacting factors for this to be practical (compression
and decompression speeds at different sparsities and message sizes,
heterogeneous link bandwidths, densification rate, etc.). Empirical tuning is a
practical choice and incurs only a one-time offline cost.

Observe that our thresholding algorithm permits one or more dense-send
ring steps to be sandwiched between sparse-send ring steps. After a dense-send
step (where the nonzero count is not computed), measured data sparsity is unavailable at the
next ring step.  This necessitates that we extrapolate densification forward
through (potentially multiple) dense-send ring steps, using Equation
\ref{fig:extrapolate_equation}, where $\sigma_{\text{prev}}$ may now represent
a prior-step computed sparsity or a previous extrapolation. When
compressions occur after extrapolations, they calibrate the tracked sparsity
using the computed nonzero count.

\subsection{COO Support in SpCCL}

In addition to Pici, we implement support for COO in our collectives and
perform additional sweeps to identify appropriate COO thresholds.  These
thresholds are higher than for Pici, at $intra\_thresh=0.8$,
$inter\_thresh=0.7$, and $ag\_thresh=0.5$, respectively.
Higher thresholds are expected because COO becomes less
space-efficient relative to Pici as sparsity decreases.

\begin{figure}[h]
  \centering
  \includegraphics[width=\columnwidth]{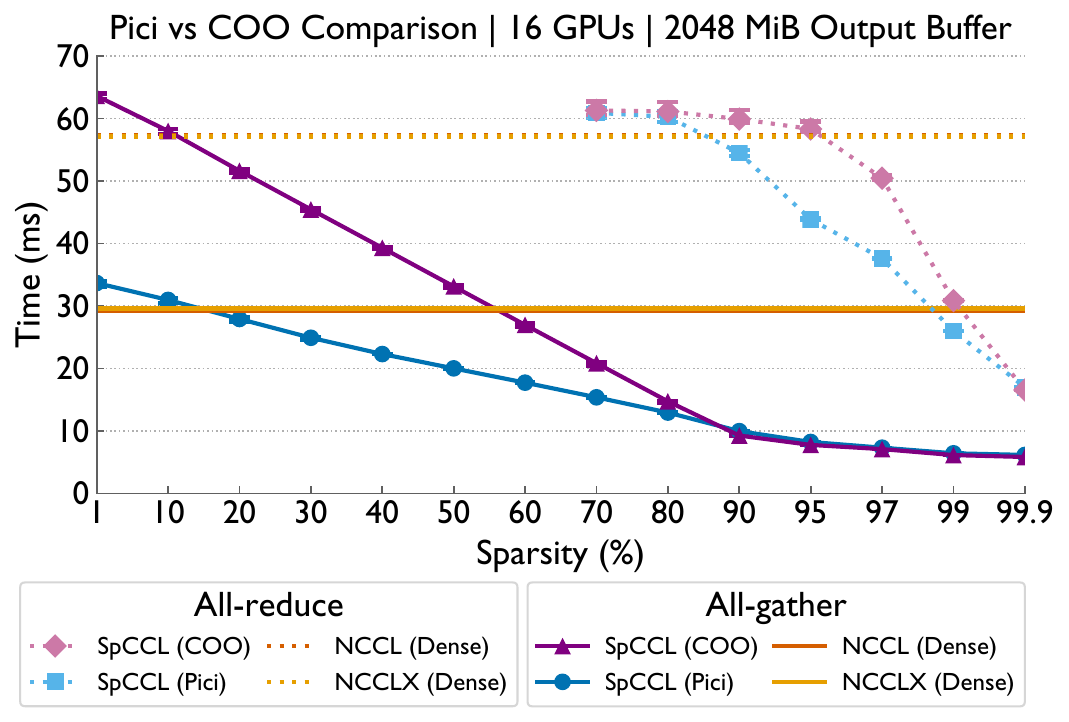}
  \caption{Comparison of Pici and COO in SpCCL with NCCL (Dense) and NCCLX (Dense) for all-gather and all-reduce on 16 GPUs.
  While COO can perform comparably or slightly better than Pici at higher sparsities, it becomes less effective as sparsity decreases, especially for all-reduce. Results are similar for other tested buffer sizes and GPU counts.}
  \label{fig:ag_ar_spop_coo_lines_fp32_16gpu_2048mib}
\end{figure}

Figure \ref{fig:ag_ar_spop_coo_lines_fp32_16gpu_2048mib} compares SpCCL using
Pici and COO with dense NCCL and NCCLX
baselines for sparse all-gather and all-reduce. While COO achieves comparable
or slightly superior performance to Pici at higher sparsities, it falls off
significantly as sparsity decreases, especially for all-reduce, where it
is comparable only within the HS regime above 99\% sparsity before both
methods approach dense baseline performance. For
all-reduce, Pici is $1.19\times$ faster than COO at 99\% sparsity and
$1.33\times$ faster at 95\% (where COO is slower than dense). For all-gather,
COO remains competitive at slightly lower sparsities (down to around 90\%) than
would be expected based on space requirements alone.  We attribute this to its
simpler decompression routine, because decompression dominates all-gather
overhead.  While compression is similar to Pici in structure (identify
nonzeros, build an index for stream compaction, compact nonzeros),
decompression only requires reading index-value pairs.

\section{Implementing Sparse Collectives in NCCLX}\label{ncclx-section}
As previously mentioned, SpCCL collectives are implemented as extensions to
NCCLX in order to exploit preexisting collective optimizations rather than
reimplementing them in a redundant engineering effort \cite{si2025collective}.
Our implementation uses NCCLX's underlying
communication components that are largely unchanged from stock NCCL, so our
descriptions of how NCCLX operates apply to NCCL as well.

\subsection{Overview of the NCCLX Collective Library}

We now provide an overview of NCCLX concepts that are relevant to our
implementation and optimizations.  For a more comprehensive analysis of NCCL,
we refer the reader to the work of Hu et al.
\cite{hu2025demystifyingncclindepthanalysis}.  Because NCCLX collectives are
implemented in CUDA kernels, they use streaming multiprocessors (SMs) to handle
GPU work. Without sufficient SMs, a collective can fail to post enough work to
saturate links. To address this, NCCLX splits collectives into
\textbf{channels}. Each channel corresponds to a single CUDA block that
executes on a single SM, and each channel has an independent progress engine
driving progress on disjoint segments of the data.

In the Simple communication protocol, after dividing the work that each process is
responsible for into channels, NCCLX further subdivides work into chunks of
\textbf{slices}.  Slices are the granularity of individual network
transmissions, and they are pipelined using send and receive FIFO buffers, with
pointers into these buffers controlling producer-consumer synchronization.  For
intra-node receives with our configuration, the receive FIFO is the peer-mapped
memory comprising a peer's send FIFO (enabled by GPUDirect P2P over NVLink
\cite{gpuDirect}).
For inter-node receives, a host-side proxy thread interacts with NICs, which
write data directly to GPU DRAM without going through the host. The ``receive +
reduce (AR/RS) + send'' slice lifecycle is handled by fused NCCLX primitives
that instantiate different versions of the templated \texttt{genericOp} device
function. Other communication protocols would complicate compression and decompression and are designed for low-latency regimes.

\subsection{Extending NCCLX to Support SpCCL Algorithms}

To support compression in NCCLX, we must decide at which granularity to apply
it. The most natural solution is to compress data at the slice level
immediately before sends are posted and to decompress immediately upon receipt.
Compressing data at any other granularity would clash with NCCLX's architecture
and interfere with pipelining, as the slice is the unit of pipelining.  As a
consequence, we implement compression and decompression within
\texttt{genericOp}.

When sending dense-format messages, the collective determines message sizes statically
based on input data size. When sending messages in a sparse format with SpCCL, the
collective cannot determine message sizes or message format in advance for any
individual receipt.
To communicate message formats and sizes to peers, we extend the Simple
protocol by prepending a 48-byte header to all messages.  The header contains
format and message size metadata, as well as offsets indicating where different
components of our sparse formats begin.

Sparse representations can exceed the size of dense representations as sparsity
decreases, so we also enlarge the FIFO buffers to ensure that even a worst-case
compressed
slice fits within its FIFO reservation. For Pici, format overhead is only
3.15\% relative to the dense representation, a significant advantage over COO's worst-case, in which
one index is needed per nonzero value.
For reduction, NCCLX accumulates directly into the send FIFO. For sparse steps,
we cannot do this, as we must compress into the send FIFO.
We reuse the application send/input buffer as accumulation scratch and then
compress into the send FIFO buffer. This allows us to avoid allocating
additional scratch space and preserves dense collective signatures.
For dense reductions, we reuse heavily vectorized NCCLX primitives such as
\texttt{reduceCopy}.

\subsection{Optimization 1: Increasing Parallelism with Channels}

\begin{figure}[h]
  \centering
  \includegraphics[width=\linewidth]{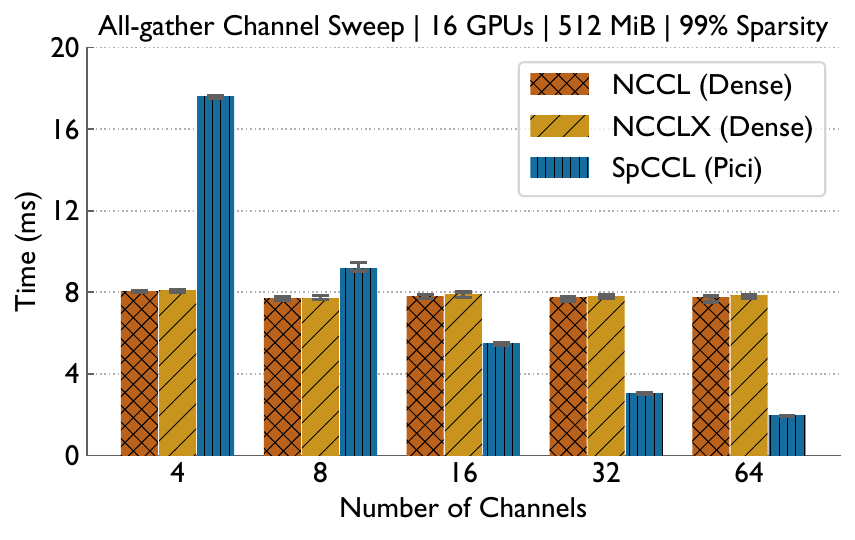}
    \caption{Impact of the number of channels used on all-gather performance on 16 GPUs (512 MiB output, 99\% sparsity).
    Observe that sparse all-gather (Pici) benefits substantially from additional channels.
    }
  \label{fig:ag_channel_sweep_32g_256mib_fp16_99}
\end{figure}

NCCLX heuristically adjusts the number of channels a collective uses (capped at
64).
When relying on these heuristics, we observe that our collectives significantly
underperform dense baselines.  We are incurring additional compute overhead for
each (generally smaller) message, so the default heuristics about how many
channels are necessary to saturate the links no longer hold. We experiment with
raising the number of channels our collectives utilize, sweeping 4–64 channels
across multiple GPU counts and input/output buffer sizes. Figure
\ref{fig:ag_channel_sweep_32g_256mib_fp16_99} illustrates the results of one of
our channel sweeps.

We observe that while dense collectives generally do not benefit from using a
large number of channels, sparse collectives benefit substantially from
elevated channel counts. For example, 64 channels is best 72.5\% of the time
for sparse all-gather (with 32 channels being best in remaining cases). Cases
where 32 channels win tend to be those with the smallest collective inputs. The
maximum size of a slice is fixed by the implementation and runtime
configuration.  Thus, with small enough inputs, adding more channels can result in
multiple channels each sending the same number of smaller slices rather than
reducing the number of per-channel slices.

Using more channels can reduce the potential for computation-communication
overlap due to the collective utilizing more SMs and placing more concurrent
demand on shared memory-system resources.  Thus, SpCCL's current implementation
may be less well-suited to workloads with heavy computation-communication
overlap.
The overlap and contention concerns may be partially mitigated in cases where
sparsity provides speedups, as the SMs remain occupied by the collective for
less time.

\subsection{Optimization 2: Avoiding Uncoalesced Reads over NVLink}

For intra-node receives, data is available via the FIFO buffer (peer-mapped
memory). For dense accumulation, the memory access pattern is perfectly
coalesced. However, scatter-adds from compressed data result in uncoalesced
reads over NVLink. As NVLink has much lower bandwidth than local HBM, when
receiving Pici-formatted data, we first copy the received data into a staging
buffer and perform the scatter-add from local memory. To avoid allocating an
additional staging buffer, we reuse the send FIFO. This is safe, as the
scatter-add must complete before we compress back into the send FIFO.

\begin{figure*}[!h]
  \centering
  \includegraphics[width=\textwidth]{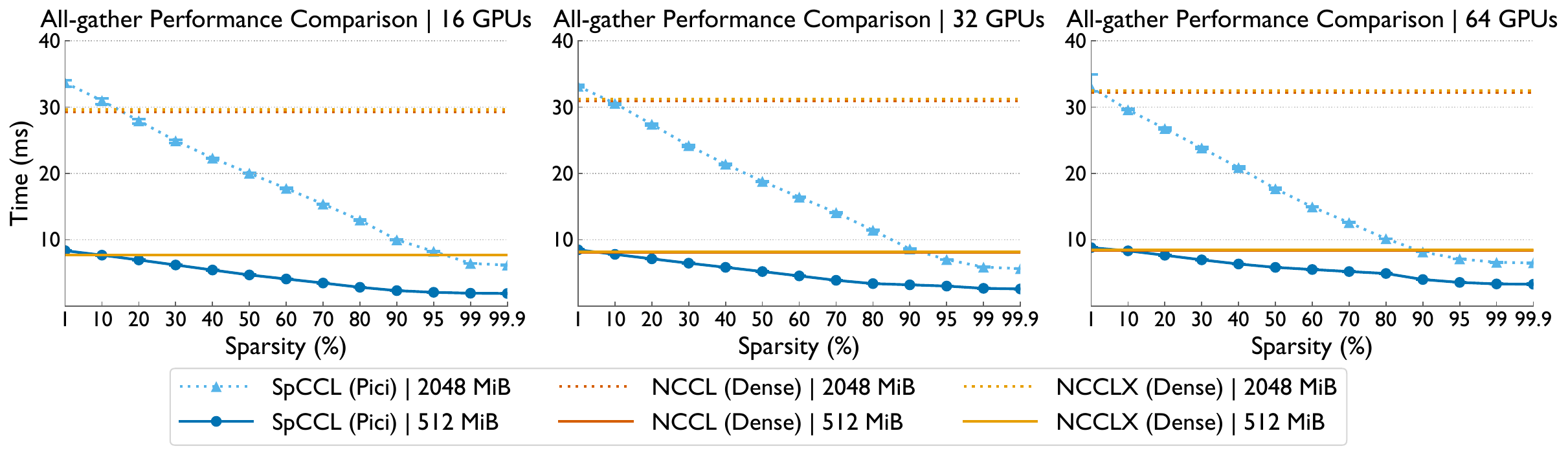}
  \caption{Performance comparison of SpCCL all-gather (Pici) versus NCCL (Dense) and NCCLX (Dense) baselines on several GPU counts for 512 and 2048 MiB output sizes. SpCCL all-gather significantly outperforms dense baselines, only falling behind dense at 1–10\% input sparsity.}
  \label{fig:ag_sparsity_panels_fp32_16_32_64_512_2048mib_output}
\end{figure*}

\section{Experimental Setup}
Below, we provide an overview of our methodology for evaluating SpCCL
collectives on Perlmutter \cite{perlmutter}.

\subsection{Collective Microbenchmarks}

First, using Pici, we compare the standalone performance of SpCCL all-gather,
reduce-scatter, and all-reduce with NCCL/NCCLX baselines. For all-gather,
we report the per-GPU output size; for reduce-scatter, the input
size; for all-reduce, the shared input/output size. We sweep a range of
sparsity levels. Each element is independently selected to be nonzero with probability
equal to $1-\sigma$. Each process uses a distinct pseudorandom seed.
With respect to densification, this is a pessimistic non-adversarial
scenario.

For all-reduce, we also compare against SparCML sparse all-reduce
implementations \cite{sparcml}. SparCML supports four algorithms for
all-reduce, including ring and recursive variants. OmniReduce is a potential
additional sparse baseline, but its reliance on InfiniBand RDMA verbs prevents
us from using it on Perlmutter without porting to CXI libfabric
\cite{omnireduce}.

We conduct all experiments on Perlmutter \cite{perlmutter}. GPU nodes have
either four 40 GB or four 80 GB A100 GPUs. We use 40 GB nodes in all
microbenchmarks. Inter-node communication occurs over the Slingshot 11
interconnect fabric, while intra-node communication uses NVLink connections.

\begin{figure*}[t]
  \centering
  \includegraphics[width=\textwidth]{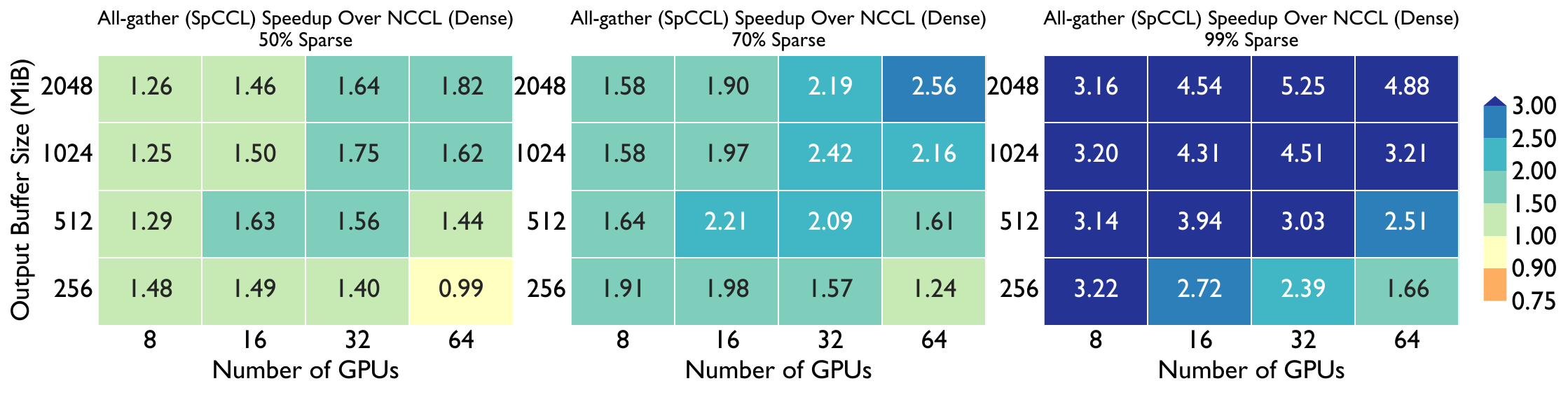}
  \caption{SpCCL all-gather (Pici) speedups over NCCL all-gather (Dense). Speedups are greatest for large output buffer sizes and at higher sparsities, but substantial speedups are achieved even at 50\% sparsity. SpCCL all-gather achieves speedups of up to $5.25\times$ at 99\% sparsity.}
  \label{fig:ag_spop_heatmap_fp32}
\end{figure*}

We compile SpCCL and NCCLX against CUDA 12.9.  Our dense NCCL baseline uses
Perlmutter's stock nccl/2.24.3 module. We obtain NCCL runtimes with NVIDIA's
nccl-tests software \cite{nccl-tests}. We implement a microbenchmarking harness
based on nccl-tests and osu-micro-benchmarks for evaluating NCCLX, SpCCL, and
SparCML (but make no modification to SparCML collectives)
\cite{panda2021mvapich,osu_micro_benchmarks}.

To account for network variability~\cite{smith:sc2018, bhatele:ipdps2020, wei:ipdps2026b}, we run each tested configuration across at
least five separate Slurm jobs. Within each non-tuning job, each configuration runs five
warmups and 1000 timed iterations.
We report runtimes as the average across jobs,
processes, and iterations at the best tested channel count, with an untimed barrier
before each iteration and a timed barrier at the end of each iteration
(consistent with nccl-tests behavior). For Pici, we set
$inter\_thresh=0.5$, $intra\_thresh=0.6$, and $ag\_thresh=0.1$. We remove
obvious outlier data points and additionally plot min/max times across jobs (never
$<4$ after filtering). SparCML runs use 50 timed iterations due to
significantly elevated runtimes relative to NCCL and SpCCL. For each SparCML
data point, we report results from the best-performing algorithm.

\subsection{Application Case Study: Gradient-Pruned DDP Training}

We evaluate SpCCL (Pici) on gradient-pruned Distributed Data Parallel (DDP)
training~\cite{pytorchdist-vldb}. We implement gradient pruning using a
sampling-based top-$k$ approach~\cite{lin2017deep}, imposing 99\% collective
input sparsity and determining the pruning threshold by sampling 0.1\% of
gradients. We further implement a leaky error feedback (EF)
mechanism~\cite{seide14_interspeech,lin2017deep} in a Triton kernel. Although
collective input sparsity is 99\%, process-dependent nonzero positions cause
sparsity to decrease during the all-reduce.

We experiment with a $1.5$B GPT-2 XL-based~\cite{gpt-2} model and a $3.3$B
StarCoder2-3B-based~\cite{lozhkov2024starcoder} model on 40 GB and 80 GB GPU
nodes, respectively. We reduce the latter to 28 layers due to the additional
memory overhead of EF. We set the sequence length to $512$ and the micro-batch
size to two per GPU. We train for $100$ iterations on the WikiText-103
dataset~\cite{wikitext-103} using Megatron-LM as our training
framework~\cite{megatronlm}. We run each configuration three times, reporting
mean runtimes of the last $80$ iterations. We discard the first $20$ iterations
as warmup. The breakdowns in Figure \ref{fig:ddp_stacked_plot} use only the run with the median mean iteration time for each configuration.

\section{Results}
We now discuss and interpret our results from benchmarking
individual collectives and from the training case study.

\subsection{Microbenchmark Results}

\begin{figure*}[t]
  \centering
  \includegraphics[width=\textwidth]{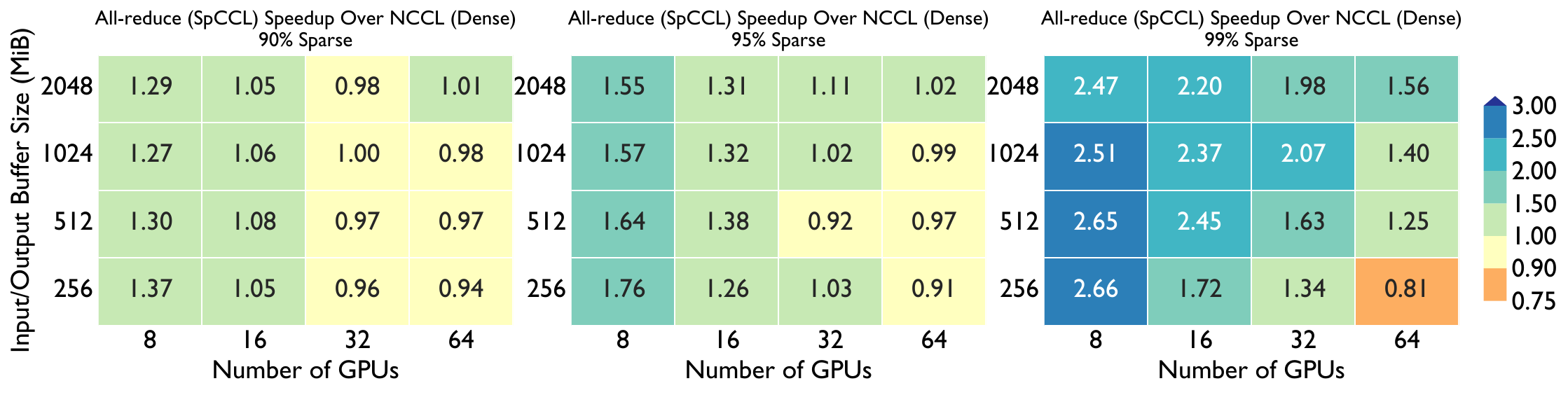}
  \caption{SpCCL all-reduce (Pici) speedups over NCCL (Dense). At 99\% sparsity, substantial
  speedups can be achieved even at 64 GPUs, while for lower sparsities of 90\% and 95\%, speedups are only
  consistently observed on lower GPU counts due to densification. SpCCL all-reduce achieves speedups of up to
  $2.66\times$ at 99\% sparsity.}
  \label{fig:ar_spop_heatmap_fp32}
\end{figure*}

\begin{figure*}[t]
  \centering
  \includegraphics[width=\textwidth]{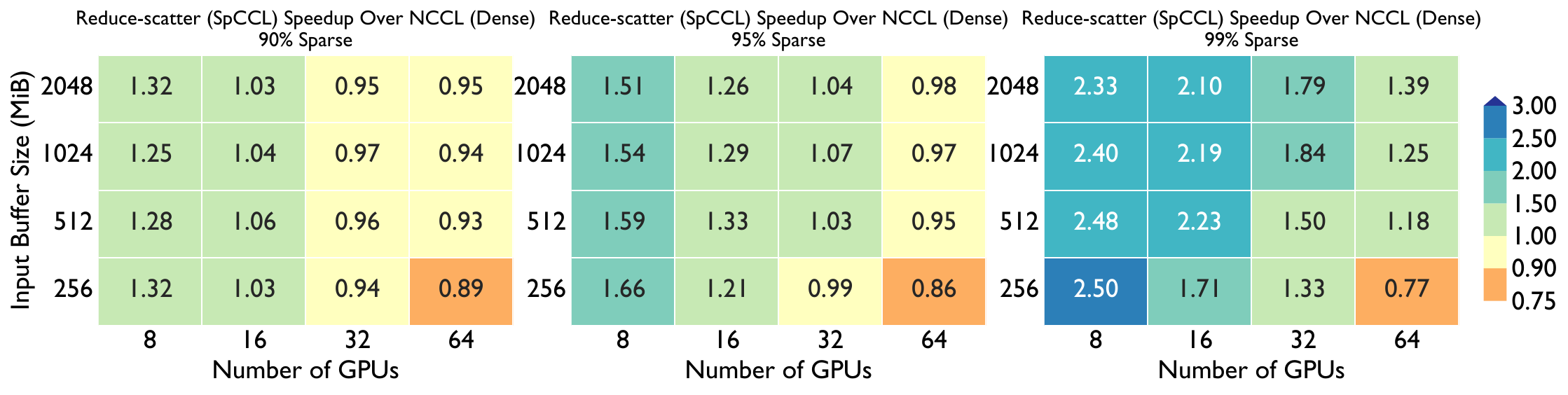}
  \caption{SpCCL reduce-scatter (Pici) speedups over NCCL (Dense). At 99\% sparsity, substantial
  speedups can be achieved even at 64 GPUs, while for lower sparsities of 90\% and 95\%, speedups are only
  consistently observed on lower GPU counts due to densification. SpCCL reduce-scatter achieves speedups of up to
  $2.5\times$ at 99\% sparsity.}
  \label{fig:rs_spop_heatmap_fp32}
\end{figure*}

\begin{figure*}[t]
  \centering
  \includegraphics[width=\textwidth]{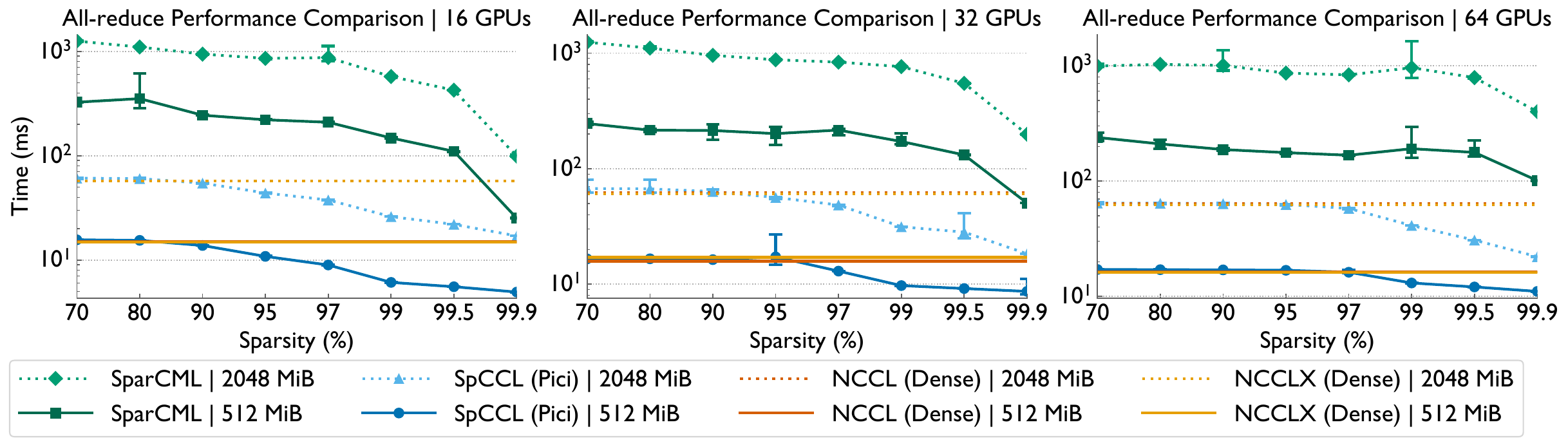}
  \caption{Comparison of SpCCL all-reduce (Pici) with dense (NCCL and NCCLX) baselines and SparCML on several GPU counts and input buffer sizes. Speedups are highest at higher input sparsities and decrease (due to densification) as sparsity decreases or as GPU counts increase.
}
  \label{fig:ar_sparsity_panels_fp32_16_32_64_512_2048mib}
\end{figure*}

NCCL and NCCLX dense baselines achieve near-identical performance (Figures
\ref{fig:ag_sparsity_panels_fp32_16_32_64_512_2048mib_output},
\ref{fig:ar_sparsity_panels_fp32_16_32_64_512_2048mib}).  Based on our cost
model, we expect to observe the largest speedups over dense baselines at the highest
sparsities, with greater speedups for all-gather than for reduce-scatter and
all-reduce. This is consistent with our results. Using Pici, SpCCL all-gather
achieves speedups over dense at sparsities as low as 10\%, highlighting Pici's
fast decompression and ability to save space even at 10\% sparsity (Figure
\ref{fig:ag_sparsity_panels_fp32_16_32_64_512_2048mib_output}, 64 GPUs, 2048
MiB).  At 99\% sparsity, SpCCL all-gather achieves speedups of up to
$5.25\times$ (Figure \ref{fig:ag_spop_heatmap_fp32}). Additionally, when we
hold per-GPU input size constant for all-gather, we observe that speedups over
NCCL actually \textit{increase} with GPU count. This can be observed in the
bottom-left-to-top-right diagonals in the
heatmaps of Figure \ref{fig:ag_spop_heatmap_fp32}
(rows refer to the output buffer size). This is likely due in part to
the step-0-only compression cost being amortized over more ring steps.

For sparse all-reduce and reduce-scatter, speedups over NCCL at a given input
sparsity decrease as process count rises (Figures
\ref{fig:ar_spop_heatmap_fp32} and \ref{fig:rs_spop_heatmap_fp32}).  This is
expected due to densification, as each additional ring step decreases sparsity.
Maximum speedups at 99\% sparsity are $2.66\times$ (AR) and $2.5\times$ (RS).
At 99\% sparsity, worthwhile speedups can be observed even up to 64 GPUs.  The
exception is the smallest 256 MiB input.

The per-process per-channel contribution at 64 GPUs with 32 active channels while
using stock NCCLX is only 128 KiB for a 256 MiB input buffer. With the default
FIFO buffer size, this is split over two slices, yielding 64 KiB individual message
sizes. Prior work establishes MPI RMA put latencies on Perlmutter over
Slingshot 11 to be about 2.49 $\mu$s \cite{hargrove2022gasnet}.  The crossover
point between latency-bound and bandwidth-bound messages in the Hockney model
is $n^* \approx \alpha/\beta$, which for Perlmutter inter-node messages is
$60.8\;\text{KiB} \approx \frac{2.49\;\mu \text{s}}{1 / 25\;\text{GB/s}}$.
Although MPI RMA put latencies are not perfectly representative of our NCCLX
send path and the exact slice size may vary by FIFO size, this is consistent
with densely formatted slices being near the latency/bandwidth crossover,
before accounting for compression and decompression overheads. Thus, the slowdown at 256 MiB
inputs on 64 GPUs is not surprising.
For such messages, our threshold choices based on the bandwidth-bound regime
are no longer reasonable, and it would be better to set very aggressive
thresholds or fall back to dense NCCLX.

Figure \ref{fig:ar_sparsity_panels_fp32_16_32_64_512_2048mib} demonstrates that
SparCML is not competitive, even against dense NCCL/NCCLX. In most cases, it is
at least an order of magnitude slower.
The SparCML authors originally report speedups against dense MPI. However, SparCML
is not GPU-aware, and the original experiments used much slower networks. Our
results highlight the utility in starting a sparse collective implementation
from a strong dense baseline on modern networks.

\begin{figure}[!h]
\centering
\includegraphics[width=\columnwidth]{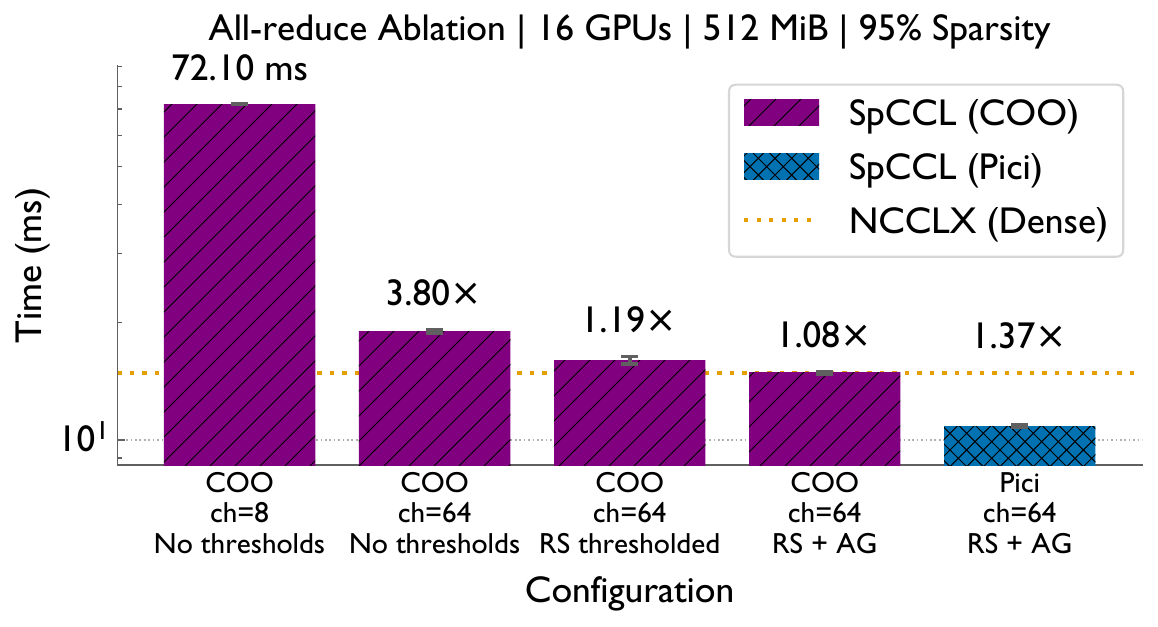}%
    \caption{Ablation demonstrating the effect of increasing channels, applying thresholds during the reduce-scatter (RS) and all-gather (AG) phases, and exchanging COO for Pici for 95\% sparse all-reduce inputs on 16 GPUs. Annotations represent speedups relative to the preceding configuration.}%
    \label{fig:ar_ablation}%
\end{figure}

Figure \ref{fig:ar_ablation} isolates the impact of SpCCL's most significant
design choices. Observe that each design decision or optimization produces a
measurable speedup, and only the fully optimized configuration outperforms
NCCLX in this experiment configuration. Increasing the channel count
provides the largest individual improvement ($3.80\times$).

\subsection{Case Study: Gradient-Pruned DDP Training}

\begin{figure}[tb]
\centering
\includegraphics[width=\columnwidth]{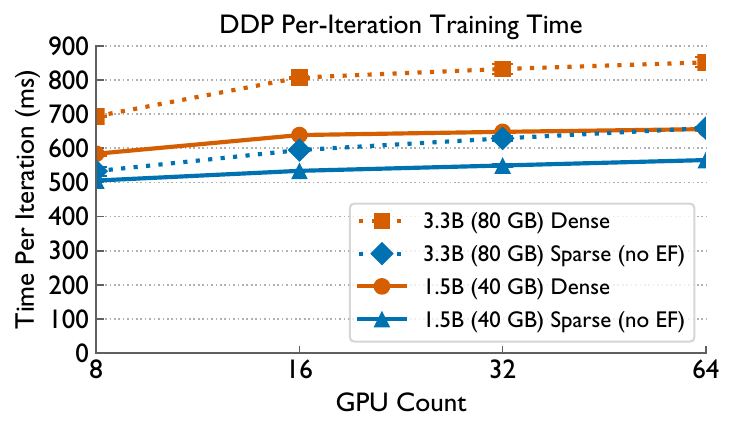}%
\caption{Performance impact of using SpCCL in DDP training for 40\,GB and 80\,GB GPUs.
    Pruning improves end-to-end time for all GPU counts.}%
\label{fig:ddp_line_plot}%
\end{figure}

Next, we evaluate SpCCL in an end-to-end gradient-pruned DDP training
application. Figure~\ref{fig:ddp_line_plot} shows weak scaling performance
across GPU counts for 40\,GB and 80\,GB GPUs. End-to-end performance with SpCCL
is at least $13\%$ faster than the dense baseline on eight 40\,GB GPUs and
$22\%$ faster on 64 80\,GB GPUs. The improvement peaks at 16 GPUs, reaching
$16\%$ and $26\%$ when running on 40\,GB and 80\,GB GPUs, respectively.

\begin{figure}[tb]
\centering
\includegraphics[width=\columnwidth]{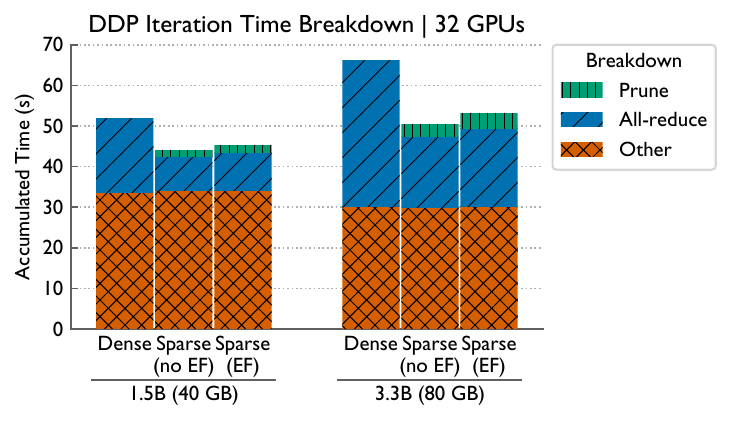}
\caption{Breakdown of all-reduce and pruning time on 32 GPUs.
    All-reduce constitutes a larger fraction of total time in 80\,GB runs than in 40\,GB runs.
    In all cases, sparse collectives reduce all-reduce time by over $40\%$.}
\label{fig:ddp_stacked_plot}
\end{figure}

Figure~\ref{fig:ddp_stacked_plot} illustrates the time spent in all-reduce and
pruning over 80 iterations on 32 GPUs with and without EF. The 80\,GB runs
spend more time in all-reduce than 40\,GB runs, which explains the speedup
differences. Without EF, in both configurations, SpCCL achieves over $2\times$ speedup on
the all-reduce portion, while pruning introduces $1.5$\,s and $3$\,s of
overhead for 40\,GB and 80\,GB runs, respectively. Pruning time scales linearly
with gradient size, as it compares each tensor element against the threshold
once and possibly discards it. Since we compute the threshold over a sampled
subset of gradients, the sorting cost is minimal.

Error feedback introduces overhead in both pruning and all-reduce. In scaling
tests, we do not observe EF overhead increasing with GPU count. During pruning,
additional error buffer memory traffic accounts for this cost. The source of
all-reduce overhead is less apparent. We hypothesize that as error accumulates,
the combined error and gradients become more uniform in magnitude during
certain iterations. This may reduce compression efficiency, degrading all-reduce
performance. SpCCL achieves significant speedups with and without EF,
demonstrating that sparse collectives can effectively accelerate gradient
synchronization in distributed training.

Our application study does not exhaust all sparsification strategies,
model parallelism schemes, or model scales, and we do not claim our sparsification
strategy is optimal. Rather, we demonstrate SpCCL's utility for sparse-training
research.


\section{Related Work}
Although prior work explores sparse communication collectives, few
systems implement exact collectives designed to work
with arbitrary sparsity patterns (particularly outside of all-reduce).
SparCML is
the work that is conceptually most similar to SpCCL. It supports multiple
sparse all-reduce implementations but operates on host buffers \cite{sparcml}.
Unlike SpCCL, it supports only a COO-style sparse representation.
OmniReduce implements sparse all-reduce as a streaming aggregation system using
a block-based sparse representation. Its microbenchmarks apply block-wise
sparsity rather than element-wise sparsity. Block-wise sparsity is highly
favorable to OmniReduce's approach \cite{omnireduce,omnireduce-experiments}.
Wijerathne et al.~describe block-based
collectives implemented in RCCL (exclusive to AMD GPUs) and
demonstrate speedups on block-wise sparse inputs \cite{wijerathne2025sparse}.
OmNICCL's sparse all-reduce relies on BlueField-2 SmartNICs for
in-network aggregation \cite{gu2024omniccl}. D-DOSA also relies on
DPUs/SmartNICs \cite{yu2025d}.

ZEN is a holistic sparse gradient synchronization system with all-reduce
semantics and load balancing, whereas SpCCL is a drop-in NCCL replacement
for supported collectives that works with GPU-resident unstructured sparse
inputs \cite{wang2025zen}. ZEN
also demonstrates speedups over baselines, albeit on lower-bandwidth networks
and with more support overlap (less fill-in) than in the SpCCL microbenchmarks, making
the results not directly comparable.

Several works implement \textit{lossy} sparse collectives, introducing sparsity
during the course of the collective.  This may be tolerable for some
applications, but lossy collectives are out of scope for this work: SpCCL
collectives are exact. O\textit{k}-Top\textit{k} and gTopKAllReduce exemplify
this line of research \cite{oktopk,shi2019distributed}.

Hecate and Hoefler et al.~describe collectives that are sparse in their
communication patterns, rather than in their communicated data \cite{qing2025hecate,hoefler2009sparse}.

Coruscant introduces a tile- and bitvector-based sparse matrix representation
that is similar to Pici \cite{coruscant}. However, the authors target
SpMM rather than communication or standalone whole-input CUDA compression
and decompression. Coruscant's index space overhead is $50\%$ relative to
the bitvector, compared to Pici's $\approx0.78\%$.

Our sparse collectives are exact with respect to their inputs.  Thus, specific
strategies for gradient pruning such as Deep Gradient Compression or
$\text{Gaussian}_k$ are orthogonal to our work \cite{lin2017deep,shi2019understanding}.
Such methods may benefit from our work so long as
they do not require additional sparsification during collectives or alter
communication schedules.

\section{Conclusion}
In this work, we introduce improvements for the collective communication of
sparse data in all-gather, all-reduce, and reduce-scatter across a range of
sparsities.  With Pici, we reduce the representation overhead of
communicating data in the non-high-sparsity regime without making assumptions about sparsity
structure. Optimized Pici kernels enable efficient compression and decompression. Through
our phase- and link-aware adaptive algorithm, we avoid compression and decompression
when their overhead is predicted to outweigh the benefits of sending less data.
By implementing SpCCL as an extension to NCCLX, we build our
library on a highly optimized foundation, and we highlight the integration and
optimization decisions that make the integration effective.  We demonstrate
substantial speedups over NCCL of up to $5.25\times$ for all-gather,
$2.5\times$ for reduce-scatter, and $2.66\times$ for all-reduce at 99\%
sparsity.  These gains translate to end-to-end performance improvements of up
to 26\% during gradient-pruned training of 1.5B and 3.3B LLMs on Perlmutter,
emphasizing the utility of our approach in a representative application.

\section*{Acknowledgments}
The authors wish to thank Siddharth Singh, Prajwal Singhania, Ashley Adames
Perez, Bahar Asgari, Donghyeon Joo, and Alan Zaoxing Liu for valuable advice
and feedback on this work. This research used
resources of the National Energy Research Scientific Computing Center (NERSC),
a U.S.~Department of Energy Office of Science User Facility, operated under
Contract No.~DE-AC02-05CH11231 using NERSC awards DDR-ERCAP0034262,
ALCC-ERCAP0034775, and DDR-ERCAP0037864.

\vspace{0.06in}
\noindent \textbf{AI usage:} Claude Code (Opus 4.X) and Codex (GPT 5.X) assisted with
implementing code for portions of the methodology
and experiments, plotting scripts, and minor text and
grammar polishing throughout the paper.
The authors developed the primary
conceptual contributions, and the core GPU kernels were predominantly
human-authored.

\IEEEtriggeratref{46}
\bibliographystyle{IEEEtran}
\bibliography{./bib/cite,./bib/pssg}

\end{document}